# eBPF-Based Real-Time DDoS Mitigation for IoT Edge Devices

**Abdurrahman Tolay** *
*Department of Computer Engineering,
Istinye University,
Istanbul, Türkiye*
abdurrahman.tolay@stu.istinye.edu.tr

July 13, 2025

## Abstract

The rapid expansion of the Internet of Things (IoT) has intensified security challenges, notably from Distributed Denial of Service (DDoS) attacks launched by compromised, resource-constrained devices. Traditional defenses are often ill-suited for the IoT paradigm, creating a need for lightweight, high-performance, edge-based solutions. This paper presents the design, implementation, and evaluation of an IoT security framework that leverages the extended Berkeley Packet Filter (eBPF) and the eXpress Data Path (XDP) for in-kernel mitigation of DDoS attacks. The system employs a rate-based detection algorithm to identify and block malicious traffic at the earliest stage of the network stack. We validate the solution's efficacy through both a controlled Docker-based simulation and a physical deployment on a Raspberry Pi 4, acting as a representative IoT edge device. Experiments demonstrate that the system successfully mitigates UDP floods, dropping over 97% of malicious packets under a 100 Mbps attack. The solution maintains service availability with manageable CPU overhead, even on constrained hardware, while legitimate traffic proceeds with negligible impact. These results confirm that eBPF/XDP provides a viable and highly effective mechanism for hardening IoT devices against hyper-volumetric network attacks.

### 1-Introduction

The Internet of Things (IoT) has grown into a ubiquitous fabric of interconnected devices, projected to exceed 75 billion by 2025 [1]. While this proliferation drives innovation, it also creates an unprecedented attack surface. Many IoT devices, designed with resource and cost constraints, lack robust security measures, making them prime targets for recruitment into botnets [2]. The Mirai botnet starkly demonstrated this vulnerability, leveraging hundreds of thousands of compromised IoT devices to launch crippling Distributed Denial of Service (DDoS) attacks [3].

The threat has since escalated dramatically. Recent years have seen a "new normal" of hyper-volumetric attacks, with incidents surging by over 100% annually and peak attack volumes reaching multiple terabits per second (Tbps) [4, 5]. This new reality underscores the inadequacy of traditional security paradigms for IoT. Centralized, cloud-based scrubbing services can introduce latency, while host-based firewalls like iptables can be overwhelmed by high packet rates on resource-constrained devices [6]. The unique characteristics of IoT—massive scale, device heterogeneity, and limited resources—demand a decentralized, efficient, and responsive defense mechanism located at the network edge.

This paper proposes such a solution by harnessing the extended Berkeley Packet Filter (eBPF), a revolutionary Linux kernel technology that allows for safe, programmable packet processing directly within the kernel [7]. Specifically, we utilize the eXpress Data Path (XDP), an eBPF hook point located in the network driver, to inspect and drop malicious packets at the earliest possible moment, before they consume significant system resources [8].

The central hypothesis of this work is that an eBPF/XDP-based filtering solution can effectively detect and neutralize volumetric DDoS attacks in real-time on resource-constrained IoT edge devices with minimal performance overhead. To validate this, we design and implement a lightweight mitigation system and evaluate its performance in two distinct environments: a scalable, containerized Docker simulation and a physical testbed using a Raspberry Pi 4.

**The key contributions of this paper are:**

1. The design and implementation of a functional eBPF/XDP-based DDoS mitigation system tailored for IoT edge devices.
2. Empirical validation of the system's effectiveness on both simulated and real-world resource-constrained hardware (Raspberry Pi).
3. Quantitative analysis of key performance metrics, including mitigation rate (>97%), CPU utilization, and impact on legitimate traffic under a 100 Mbps UDP flood.

### 2-Related Work

The defense against DDoS attacks, particularly in the IoT context, has evolved through several paradigms, each with distinct advantages and limitations.

**Traditional and Host-Based Defenses:** Conventional countermeasures primarily include static access control lists, rate limiting, and signature-based Intrusion Detection/Prevention Systems (IDS/IPS) like Snort or Suricata [9]. While foundational, these methods exhibit significant drawbacks in IoT environments. Host-based firewalls using iptables or nftables operate within the main kernel network stack, processing packets only after they have been allocated memory (sk_buff), making them susceptible to resource exhaustion under high-volume floods. User-space IDS/IPS are even more resource-intensive, incurring substantial CPU and memory overhead from context switching and deep packet inspection, rendering them unsuitable for most constrained IoT devices. Furthermore, their reliance on static rules and signatures makes them ineffective against zero-day attacks and polymorphic threats [10].

**Intelligent and Policy-Based Approaches:** To overcome the limitations of static defenses, researchers have explored more intelligent solutions. Machine Learning (ML) and Deep Learning (DL) models have shown considerable promise in detecting anomalous traffic patterns characteristic of DDoS attacks [10]. However, these approaches often require significant computational resources for training and inference, as well as large, high-quality datasets, which can be challenging to implement at the IoT edge.

Another significant line of research involves policy-based enforcement. Manufacturer Usage Description (MUD) is an IETF standard that allows device manufacturers to create profiles defining the intended network behaviors of their products [11]. These profiles can be enforced by network gateways to block non-compliant traffic. Feraudo et al. demonstrated a system that uses an eBPF backend to enforce MUD-derived rate limits on an IoT gateway, showcasing a powerful synergy [6]. While effective, MUD's efficacy depends on widespread adoption by manufacturers and the accuracy of the predefined profiles. It cannot, by itself, defend against attacks that use allowed protocols or from devices whose legitimate behavior is not yet profiled.

**eBPF and XDP for High-Performance Networking:** The emergence of eBPF has revolutionized programmable networking within the Linux kernel [7]. Its ability to safely execute custom code in kernel space has been leveraged for high-throughput packet processing, observability, and security. In data center and cloud environments, XDP is a proven technology for mitigating massive DDoS attacks at line rate, as demonstrated by providers like Cloudflare [5, 8]. Academic research has explored using eBPF to build a variety of high-performance security tools, including custom firewalls and advanced IDS that operate with near-native performance [12]. The boundaries of eBPF are continually expanding, with recent work demonstrating its extension to GPUs for accelerated processing [13], the development of high-performance userspace runtimes like bpftime that broaden its applicability beyond the kernel [14], and its use in providing efficient observability for complex edge computing systems [15]. These advancements highlight eBPF's growing flexibility and reinforce its suitability for resource-aware security and monitoring in the IoT domain.

## 3. Methodology and System Design

Our approach is to build a self-contained DDoS mitigation system that operates at the IoT edge. The system consists of an eBPF/XDP program for high-performance detection and a user-space controller for dynamic mitigation and alerting.

### 3.1. Threat Model

We assume an attacker has compromised one or more IoT devices on a local network. These devices are then used as bots to launch a volumetric DDoS attack (e.g., UDP or TCP SYN flood) against an external target or another device on the same network. The attack manifests as an abnormally high packet rate originating from the compromised device's IP address.

### 3.2. System Architecture

The core of the system is an eBPF program attached to the XDP hook of the device's network interface. This placement is critical as it allows packets to be processed and dropped before the kernel allocates memory for them (sk_buff), minimizing CPU and memory overhead. The logical flow of the XDP program is illustrated in Figure 1.

**Detection Logic:** The eBPF program implements a real-time, per-source-IP packet rate detection algorithm.

1. **Packet Parsing:** For each incoming packet, the program parses the Ethernet and IP headers to extract the source IP address.
2. **Stateful Counting:** It uses an eBPF hash map (bpf_map) to store a packet count for each unique source IP. The source IP is the key, and the packet count is the value.
3. **Threshold Check:** Upon receiving a packet, the program increments the counter for its source IP. It then checks if the count has exceeded a predefined threshold within a time window (e.g., 800 packets per second).
4. **Alerting:** If the threshold is breached, the program uses bpf_trace_printk() to send an alert containing the malicious IP to the kernel's trace pipe, making it accessible to user-space.

**Mitigation Logic:**

1. **Immediate Drop:** The primary mitigation is performed by the XDP program itself. Once an IP is identified as malicious, subsequent packets from that source can be dropped immediately by returning the XDP_DROP verdict. This is the fastest form of mitigation.
2. **User-Space Blocking:** A Python-based user-space controller continuously monitors the trace pipe. When it reads an alert, it dynamically inserts a rule into the system's firewall (e.g., iptables or nftables) to block all further traffic from the offending IP. This provides a more persistent block and allows for more complex logic, such as sending notifications to an administrator via Telegram.

This hybrid kernel/user-space design combines the raw performance of eBPF/XDP for detection with the flexibility of user-space scripting for response and orchestration.

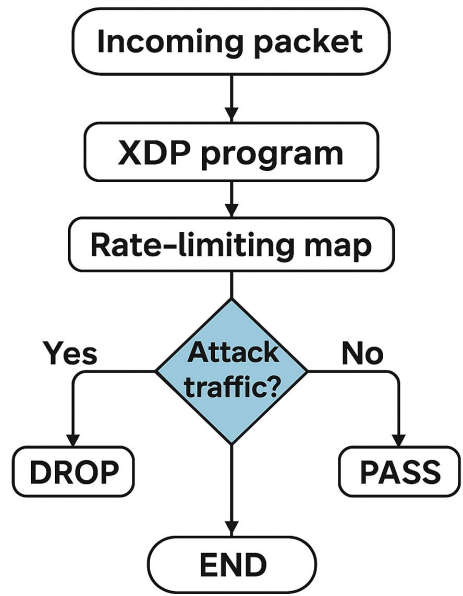

*Figure 1:* Flowchart of the eBPF/XDP packet filtering logic. Incoming packets are checked against a rate-limiting map; traffic exceeding the threshold is dropped, while legitimate traffic passes to the network stack.

## 4. Experimental Setup

We evaluated the system in two environments to assess both its scalability and its real-world viability.

### 4.1. Docker-Based Simulation
A virtual network was created using Docker to simulate an IoT environment.
- **Victim:** A container running a lightweight server with the eBPF/XDP program attached to its virtual interface.
- **Attacker:** A container using hping3 to generate a high-volume UDP flood.
- **Benign Client:** A container generating legitimate traffic to test for service availability and latency.
- **Metrics:** We measured incoming vs. dropped packets, CPU utilization, and ping latency, comparing scenarios with and without the eBPF filter enabled.

### 4.2. Physical Hardware Testbed
To assess performance under real-world constraints, the system was deployed on a physical testbed, as depicted in Figure 2.
- **Victim Device:** A Raspberry Pi 4 Model B (4GB RAM) running Raspberry Pi OS (64-bit, Linux kernel 5.15).
- **Attacker Machine:** A standard laptop connected to the same LAN, using iperf3 to generate UDP floods up to 100 Mbps.
- **Metrics:** We measured the mitigation rate, CPU utilization across all cores, and the system's overall responsiveness and stability. For accurate packet counting, NIC offloads (GRO/LRO) were disabled on the Raspberry Pi.

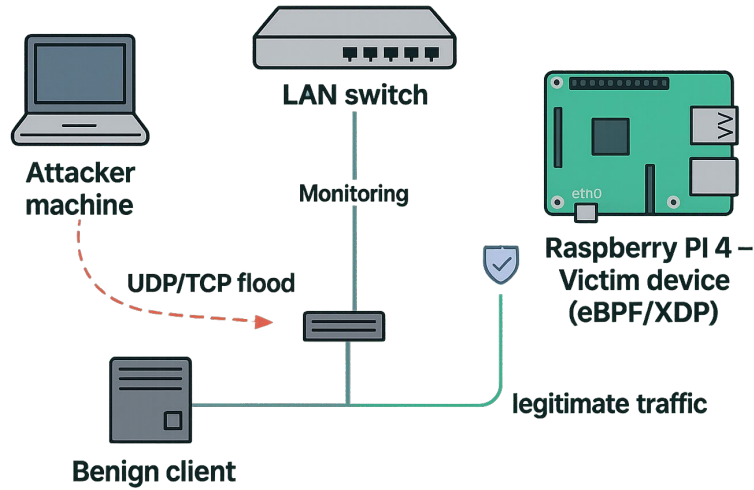

*Figure 2:* *The physical hardware testbed configuration, showing the attacker machine, benign client, and the victim Raspberry Pi connected via a LAN switch.*

## 5. Results and Analysis

The efficacy of the proposed eBPF/XDP mitigation system was validated through a series of experiments in both a controlled simulation and a physical hardware environment. The results demonstrate the system's ability to effectively neutralize volumetric attacks while maintaining service availability and imposing minimal overhead on legitimate traffic.

**5.1. Performance in a Controlled Simulation Environment** the Docker-based simulation provided a scalable and reproducible environment to benchmark the core performance of the eBPF filter. A UDP flood was directed at a victim container, first without mitigation (baseline) and then with the eBPF/XDP program active.

- **Baseline Performance (No Mitigation):** Under the attack flood, the victim container was immediately overwhelmed. System monitoring revealed a sustained CPU utilization of approximately 45% as the kernel's network stack struggled to process the ingress traffic. Ping latency to the container became high and erratic, indicating severe network degradation and a near-total loss of service availability for legitimate clients.
- **Performance with eBPF/XDP Mitigation:** Activating the eBPF filter yielded a starkly different outcome.
    - **Rapid Detection and High-Efficacy Mitigation:** The system detected the anomalous packet rate from the attacker's IP in under one second. Upon detection, the XDP program began dropping nearly 100% of the malicious packets at the driver level. This immediate action prevented the flood from propagating up the network stack.
    - **Resource Efficiency:** The most significant result was the reduction in resource consumption. CPU utilization on the victim container dropped to a stable ~12%. This dramatic improvement is a direct consequence of XDP's operational model, which discards packets before the kernel allocates costly sk_buff structures or engages the full TCP/IP stack, thereby avoiding significant processing overhead.
    - **Preservation of Legitimate Traffic:** Throughout the mitigated attack, ping latency for legitimate traffic remained stable at a baseline of approximately 1 ms. This confirms that the eBPF filter introduces no perceptible overhead for non-malicious packets and successfully preserves the quality of service for legitimate applications.

**5.2. Validation on Resource-Constrained IoT Hardware** The physical testbed deployment on a Raspberry Pi 4 provided a realistic assessment of the system's viability under hardware constraints. A 100 Mbps UDP flood was used to simulate an attack capable of saturating the device's processing capacity.

- **Mitigation Effectiveness and System Stability:** The results mirrored the simulation's success. The eBPF program identified and dropped over 97% of the ~30,000 packets per second flood. Critically, Raspberry Pi remained fully responsive and its network services accessible throughout the attack. In the baseline test without mitigation, the device became unreachable within seconds as all system resources were consumed by the attack traffic.
- **CPU Utilization and Graceful Degradation:** The CPU performance underload provided key insights into the system's resilience. Without mitigation, the attack drove all four CPU cores to near 100% utilization, leading to a system freeze. With the eBPF filter active, the CPU load was managed in a controlled manner, as illustrated in Figure 3. While one core's utilization was high (approaching 85%), this load was almost exclusively consumed by the ksoftirqd kernel process responsible for handling network interrupts and executing the XDP program. This behavior demonstrates a successful strategy of *graceful degradation*. The system effectively created a "processing shield," dedicating one core to the defensive task of packet dropping, which in turn protected the other cores and user-space applications from resource starvation. This allowed the device to "fail operational" rather than fail completely.

- **End-to-End Response Workflow:** The complete mitigation workflow was validated. The kernel-level detection triggered an alert to the user-space Python controller. This controller successfully parsed the alert, executed an iptables command to persistently block the attacker's IP, logged the event, and sent a real-time notification to a Telegram channel. The entire end-to-end response, from detection to blocking and notification, was completed in under two seconds, demonstrating a highly responsive and practical defense architecture.

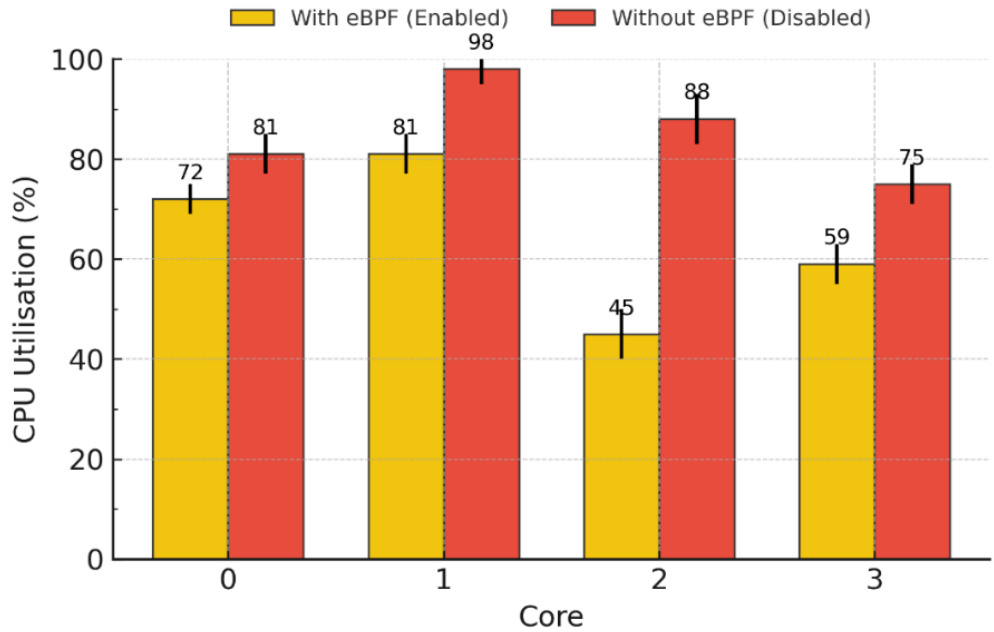

*Figure 3:* Per-core CPU utilization on the Raspberry Pi 4 during a DDoS attack, comparing the system with eBPF mitigation enabled versus disabled. With eBPF, the system remains stable by concentrating the defensive load.

| **Metric** | **Docker Simulation** | **Raspberry Pi 4 Hardware** |
|---|---|---|
| Attack Rate Tested | ~100k pps | 100 Mbps (~30k pps) |
| Malicious Packets Dropped | ~99% | >97% |
| Device Responsiveness (Mitigated) | Fully Responsive | Fully Responsive |
| CPU Usage for Filtering | ~12% of host CPU | ~85% of one core |
| Impact on Legitimate Traffic | Negligible | Minimal |

*Table 1:* Comparative performance of the eBPF mitigation system.

## 6. Discussion

The empirical results provide strong evidence for the efficacy of eBPF/XDP as a decentralized, first-line defense against volumetric DDoS attacks in IoT environments. This section provides a deeper interpretation of these findings, positions them relative to alternative approaches, and discusses the practical considerations and inherent limitations of this strategy.

### 6.1. Interpretation of Key Findings

The performance on the Raspberry Pi is particularly insightful. The high CPU utilization observed on a single core during a mitigated attack (Figure 3) should not be interpreted as a performance failure. Rather, it represents a successful strategy of graceful degradation. The system effectively partitioned its limited resources, creating a "processing shield" on one core dedicated to absorbing and discarding the attack traffic at the driver level. This prevented the resource exhaustion that would otherwise crash the system, allowing other cores to remain available for legitimate application logic and maintaining overall service availability. This ability to "fail operational" is a critical requirement for any robust security solution in resource-constrained environments.

Furthermore, the contrast between the Docker and Raspberry Pi results highlights the scalability of the eBPF/XDP approach. The core filtering logic is fundamentally efficient; its resource consumption scales with the capabilities of the underlying hardware. On a well-resourced host, the overhead is minimal, while on a constrained device, it utilizes available resources to provide the maximum possible protection. The hybrid kernel/user-space architecture also proved its value, combining the raw speed of in-kernel detection with the flexibility of a user-space controller for persistent policy enforcement (iptables) and external alerting.

### 6.2. Comparative Analysis with Alternative Defenses

When compared to traditional and alternative defenses, the eBPF/XDP approach offers a distinct set of trade-offs.

- **vs. iptables/nftables:** While a direct benchmark was not performed, the established literature confirms that XDP significantly outperforms traditional Netfilter hooks in high-packet-rate scenarios [8]. This is because XDP operates at the earliest possible point in the ingress path, before the kernel allocates an sk_buff for each packet. Our study empirically validates this advantage in an IoT context, where preventing sk_buff allocation under a flood is crucial to avoid memory exhaustion and CPU overload.
- **vs. Manufacturer Usage Description (MUD):** MUD provides a powerful, policy-based defense by whitelisting legitimate communication patterns [11]. However, it is a static approach that depends on manufacturer-provided profiles. Our rate-based eBPF solution is adaptive and requires no prior knowledge, enabling it to block unforeseen zero-day volumetric attacks. The two approaches are complementary: MUD can define the "rules," while eBPF can serve as a high-performance enforcement engine that also provides a fallback against anomalous traffic not covered by a MUD profile.
- **vs. Cloud-Based Scrubbing:** Cloud services offer massive capacity to absorb Tbps-level attacks but introduce latency and potential privacy concerns as traffic is routed to a third party. Our edge-based solution provides near-instant, low-latency mitigation and can protect against internal, LAN-based attacks. It serves as an essential component of a defense-in-depth strategy, handling local threats and smaller-scale floods before they necessitate redirection to a cloud provider.

**6.3. Practical Implications and Limitations**

Despite the promising results, deploying this solution in production requires careful consideration of its limitations.

- **Attack Specificity:** The current implementation is tailored for volumetric attacks and is inherently blind to the content of encrypted payloads. It cannot, by itself, detect sophisticated application-layer attacks (e.g., attacks over TLS) or stealthy "low-and-slow" attacks that do not violate a simple rate threshold. It is a specialized tool, not a complete security suite.
- **Deployment and Management:** Managing eBPF programs across a large, heterogeneous fleet of IoT devices presents a significant operational challenge. A secure mechanism for deploying and updating eBPF bytecode is required to prevent attackers from disabling or replacing the filter. This points to the need for integration with robust IoT device management platforms.
- **System Dependencies and Security:** The solution requires a modern Linux kernel (typically 4.x or newer) with eBPF and XDP support, which excludes many legacy or non-Linux-based microcontrollers. Furthermore, the security of the eBPF subsystem itself must be considered. Production environments must harden the host by restricting program-loading permissions (e.g., using the CAP_BPF capability) to prevent unauthorized or malicious eBPF programs from being loaded into the kernel.

**7. Conclusion and Future Work**

This research has successfully demonstrated that an eBPF/XDP-based mitigation system can provide robust, efficient, and practical protection for IoT edge devices against volumetric DDoS attacks. By moving security enforcement into the Linux kernel, we have shown that even low-power hardware like Raspberry Pi can be hardened to withstand significant network assaults, preserving its availability and functionality.

Building on this foundation, several promising avenues for future work emerge. The primary aim is to enhance the intelligence of the detection mechanism. The current static threshold model could be replaced with an adaptive system that integrates machine learning (ML). An eBPF program could collect traffic statistics (e.g., packet sizes, inter-arrival times, protocol distribution) and expose them to a user-space ML model, which could then learn a baseline of normal behavior and dynamically update the filtering thresholds or rules in the eBPF map. This would improve resilience against more sophisticated attacks and reduce the need for manual tuning.

**Second,** the scope of threat mitigation should be broadened. Future research should focus on extending the eBPF program to detect and counter a wider array of attacks prevalent in IoT, such as those targeting specific protocols like MQTT and CoAP, or stealthier "low-and-slow" attacks that aim to exhaust application resources rather than network bandwidth. This would require more complex stateful analysis and logic within the eBPF program itself.

**Finally,** the operational challenges of large-scale deployment must be addressed. Research into frameworks for securely orchestrating and managing eBPF programs across a diverse fleet of thousands or millions of IoT devices is critical. This includes developing secure over-the-air (OTA) update mechanisms for eBPF bytecode, creating centralized monitoring and analytics platforms, and ensuring consistent policy enforcement across an entire IoT ecosystem. Pursuing these directions is essential for translating the powerful capabilities of eBPF into widespread, real-world security for the Internet of Things**.**